\newif\ifarxiv
\arxivtrue

\ifarxiv
    \documentclass[11pt]{article}
    \usepackage{float}
    \usepackage{fullpage}
    \usepackage[round]{natbib}
    \usepackage{hyperref}
    \usepackage{palatino}
    \usepackage{setspace}
\else
    \documentclass[preprints,article,accept,moreauthors,pdftex]{Definitions/mdpi}
\fi

\newif\ifdraftcolors
\ifdraftcolors

\else

\fi

\usepackage[export]{adjustbox}
\usepackage{amsfonts}
\usepackage{amsmath}
\usepackage{amssymb}
\usepackage{authblk}
\usepackage[greek,english]{babel}
\usepackage{blkarray}
\usepackage{todonotes}
\usepackage{xspace}
\usepackage[braket, qm]{qcircuit}
\usepackage{graphicx}
\usepackage{tikz}
\newcommand{\andthen}{\ \&\mathrm{then}\ }

\usepackage{subfig}

\graphicspath{ {./images/} }

\ifarxiv
    
\begin{document}
    \title{Quantum Circuit Components for Cognitive Decision-Making}
    \author{Dominic Widdows}
    \affil{IonQ, Inc.} 
    \author{Jyoti Rani} 
    \affil{University of California, Berkeley} 
    \author{Emmanuel M. Pothos}
    \affil{City, University of London}
    %\affil{\normalsize Planned for \url{https://www.mdpi.com/journal/entropy/special_issues/SWMF66G100}}
    %\affil{\bf{Please do not share until this appears on arxiv.}}
    %\affil{\normalsize{Overleaf URL: \url{https://www.overleaf.com/project/62cc89747638f74a754846ff}}}
    %\date{\today}
    \maketitle
\else
    \firstpage{1} 
\makeatletter 
\setcounter{page}{\@firstpage} 
\makeatother
\pubvolume{1}
\issuenum{1}
\articlenumber{0}
\pubyear{2023}
\copyrightyear{2022}
%\externaleditor{Academic Editor: Firstname Lastname}
\datereceived{} 
\daterevised{}
\dateaccepted{} 
\datepublished{} 
%\datecorrected{} % Corrected papers include a "Corrected: XXX" date in the original paper.
%\dateretracted{} % Corrected papers include a "Retracted: XXX" date in the original paper.
\hreflink{https://doi.org/} % If needed use \linebreak
%\doinum{}
%------------------------------------------------------------------
% The following line should be uncommented if the LaTeX file is uploaded to arXiv.org
%\pdfoutput=1

%=================================================================
% Add packages and commands here. The following packages are loaded in our class file: fontenc, inputenc, calc, indentfirst, fancyhdr, graphicx, epstopdf, lastpage, ifthen, lineno, float, amsmath, amssymb, setspace, enumitem, mathpazo, booktabs, titlesec, etoolbox, tabto, xcolor, colortbl, soul, multirow, microtype, tikz, totcount, changepage, attrib, upgreek, array, tabularx, pbox, ragged2e, tocloft, marginnote, marginfix, enotez, amsthm, natbib, hyperref, scrextend, url, geometry, newfloat, caption, draftwatermark, seqsplit

%=================================================================
%% Please use the following mathematics environments: Theorem, Lemma, Corollary, Proposition, Characterization, Property, Problem, Example, ExamplesandDefinitions, Hypothesis, Remark, Definition, Notation, Assumption
%% For proofs, please use the proof environment (the amsthm package is loaded by the MDPI class).

%=================================================================
% Full title of the paper (Capitalized)
\Title{Quantum Circuit Components for Cognitive Decision-Making}

% \author{Jyoti Rani, Dominic Widdows, maybe +friends}
% \affil{IonQ, Inc.} 

% MDPI internal command: Title for citation in the left column
\TitleCitation{Title}

% Author Orchid ID: enter ID or remove command
%\newcommand{\orcidauthorA}{0000-0000-0000-000X} % Add \orcidA{} behind the author's name
%\newcommand{\orcidauthorB}{0000-0000-0000-000X} % Add \orcidB{} behind the author's name

% Authors, for the paper (add full first names)
\Author{Dominic Widdows $^{1}$, Jyoti Rani $^{2}$, and Emmanuel Pothos $^{3}$}

%\longauthorlist{yes}

% MDPI internal command: Authors, for metadata in PDF
\AuthorNames{Dominic Widdows, Jyoti Rani, Emmanuel Pothos}

% MDPI internal command: Authors, for citation in the left column
%\AuthorCitation{Lastname, F.; Lastname, F.; Lastname, F.}
% If this is a Chicago style journal: Lastname, Firstname, Firstname Lastname, and Firstname Lastname.

% Affiliations / Addresses (Add [1] after \address if there is only one affiliation.)
\address{%
$^{1}$ \quad IonQ, Inc; widdows@ionq.com \\
$^{2}$ \quad University of California, Berkeley; jyoti.rani@berkeley.edu \\
$^{3}$ \quad City, University of London; emmanuel.pothos.1@city.ac.uk
}

% Contact information of the corresponding author
% \corres{Correspondence: e-mail@e-mail.com; Tel.: (optional; include country code; if there are multiple corresponding authors, add author initials) +xx-xxxx-xxx-xxxx (F.L.)}

\fi

\centering
Preprint of paper published in {\it Entropy} 2023, Volume 25, Issue 4, 548.
\url{https://www.mdpi.com/1099-4300/24/11/1592}

Please cite the version published in {\it Entropy}.

\abstract{This paper demonstrates that some non-classical models of human decision-making
	can be run successfully as circuits on quantum computers.
	Since the 1960s, many observed cognitive behaviors have been shown to violate rules 
	based on classical probability and set theory. 
	For example, the order in which questions are posed in a survey affects whether participants answer `yes' or `no',
	so the population that answers `yes' to both questions cannot be modeled as the intersection of two fixed sets.
	It can, however, be modeled as a sequence of projections carried out in different orders. This and other examples
	have been described successfully using quantum probability, which relies on comparing angles between
	subspaces rather than volumes between subsets.
	Now in the early 2020s, quantum computers have reached the point where some of these quantum cognitive
	models can be implemented and investigated on quantum hardware, by representing the mental states in 
	qubit registers, and the cognitive operations and decisions using different gates and measurements.
	This paper develops such quantum circuit representations for quantum cognitive models,
	focusing particularly on modeling order effects and decision-making under uncertainty. 
	The claim is not that the human brain uses qubits and quantum circuits explicitly
	(just like the use of Boolean set theory does not require the brain to be using classical bits),
	but that the mathematics shared between quantum cognition and quantum computing
	motivates the exploration of quantum computers for cognition modeling.
	Key quantum properties include superposition, entanglement, and collapse, as these 
	mathematical elements provide a common language between cognitive models, quantum
	hardware, and circuit implementations.}

%\keyword{\highlighting{Cognitive Decision-Making, Quantum Cognition, Quantum Computing} %MDPI: Please list three to ten pertinent keywords specific to the article; yet reasonably common within the subject discipline. %EditedHere
%} 	
	
	%%%%%%%%%
	\section{Introduction}
	
	Human behavior often evades predictions made by mathematical laws, and 
	models based on classical mechanics and probability have sometimes proved disappointing \citep{blutner2016quantum}. 
	Much of the information we depend on is open to doubt, almost all of the predictions we make using this information
	are uncertain \citep{pothos2022quantum}, 
	and even models based on the assumption that the future
	will be like the past, even in a statistical sense, have sometimes been quite inadequate \citep{cady2015what} (Ch 8). 
	Events and beliefs can be connected in ways we do not always understand, and we cannot always explore
	such situations without disturbing them.
	
	In light of these problems, some traditional objections to quantum theory look like virtues.
	Unpredictability is part of the universe. Sometimes observing one thing makes another harder to observe.
	Asking a question can itself force a decision that closes other options. Once someone has chosen a position,
	even arbitrarily, they are likely to stick to it unless the situation evolves.
	
	Motivating analogies alone are not evidence that quantum theory or computing are better at modeling or implementing
	human-like behavior. Such evidence has been carefully collected in works including those 
	of \citet{aerts2009quantum,busemeyer2012quantum,blutner2016quantum}.
	The correspondences are mathematical rather than physical, involving vectors, bases, matrix operations,
	and projections,
	rather than waves, particles, spin, and angular momentum.
	Quantum cognitive models do not depend on their implementation using quantum mechanics at the atomic level,
	any more than the use of calculus in classical economics depends on the ability to trade infinitesimal
	amounts of money.
	
	Now in 2023, we have working quantum computers as well. These machines are at an early 
	rapid-growth stage:
	for example, the machine used in the experiments reported here uses 11 qubits \citep{wright2019benchmarking},
	and the number of available and reliable qubits is already in the twenties, which can support larger and deeper circuits \citep{ionq2022aria}.
	The time is ripe to investigate what these machines are capable of, especially in fields where
	quantum mathematical concepts have already proved useful. Such approaches are being tried
	in natural language processing~\citep{lorenz2021qnlp,widdows2022near} and economics \citep{orrell2020quantum}.
	
	In this paper, we show some initial quantum circuit implementations
	that tie the cognitive and mathematical modeling together into intuitive building-blocks. 
	The models are taken from explicit small-scale examples in the psychology literature, so the 
	implementations described can all be comfortably simulated on a modest classical computer---the results here do not claim to show 'quantum advantage' in a computational sense. 
	Instead, we show that very small circuits---some as small as a single qubit---can model surprisingly human scenarios.
	In addition, it is sometimes easy to compose or wrap these components: for example, using entanglement
	with one extra qubit can activate different behaviors for different inputs. 
	
	The paper proceeds by reviewing some long-established violations of classical set-theoretic reasoning (Section \ref{sec:violations}),
	and successful alternatives based on vectors rather than sets developed by quantum cognition researchers (Section \ref{sec:qc_intro}).
	Section \ref{sec:qcomp_intro} introduces quantum circuits, with enough detail to help understand the quantum circuits
	for order and disjunction effects which are developed in Sections \ref{sec:order_circuits}--\ref{sec:disjunction_circuits}.
	Section \ref{sec:hardware-sec} describes the implementation and results on quantum hardware, and Section \ref{sec:related}
	discusses related and further work.
	
	%%%%%%%%%
	\section{Human Violations of Classical Probability}
	\label{sec:violations}
	
	Everyday human judgments and choices sometimes appear to defy classical probability rules.
	For example, the order in which we ask questions matters, in ways that violate the
	classical notion that a conjunction is modeled by an intersection of fixed sets. We begin 
	by explaining this using a well-known example, the Clinton--Gore order experiment described by \citet{moore2002measuring}. 
	
	In the 1990s, Bill Clinton was the President 
	of the United States, and Al Gore was the Vice President. Al Gore was perceived in public 
	opinion to be “more trustworthy” than Bill Clinton (see below), but the perceptions of trustworthiness varied 
	depending on whether the two people were considered separately or together. This has been discussed from a quantum point 
	of view by several authors including \citet{wang2013quantum} who describe the results as follows:
	
	%\vspace{0.1in}
	\begin{quote}
		In a Gallup poll conducted on 6--7 September 1997, half of the 1002 respondents were asked the following pair of questions: “Do you generally think Bill Clinton is honest and trustworthy?” and subsequently, the same question about Al Gore. The other half of the respondents answered exactly the same questions but in the opposite order. ... The results of the poll exhibited a striking order effect. In the non-comparative context, Clinton received a 50\% agreement rate and Gore received 68\%, which shows a large gap of 18\%. However, in the comparative context, the agreement rate for Clinton increased to 57\% while for Gore, it decreased to 60\%.
	\end{quote}
	%\vspace{0.1in}
	
	\noindent
	{Here,}  
	the non-comparative context means asking “Is X trustworthy?” on its own, 
	whereas the comparative context
	means asking “Is Y trustworthy?” first, followed by “Is X trustworthy?”
	
	These findings are not surprising in day-to-day life, and, indeed, they are entirely consistent with social psychology theory (e.g.,
	\citep{schwarz2007attitude}). We expect that the order in which information is presented affects the way it is perceived---for example, if you had decided to ask for a pay-rise, you would give your manager good news rather than bad news just beforehand. Perhaps more surprising is the fact that order effects {\it {are not}  
	} part of classical logic or probability, unless we add new rules that introduce different conditions being
	applied for observations made in different orders. 
	It is important to understand why this is the case.
	
	The change in outcomes, depending on which order the questions are asked, violates the classical assumption
	that probability is a measure of the relative sizes of different fixed sets. For example, it
	would appear from the experiment quoted above that the size of the set $C$ of people who believe that “Clinton
	is honest” changes if people are asked “Do you believe that Gore is honest?” just beforehand.
	Classical set theory models the answers to the Clinton--Gore questions by assuming that there is a set $P$ corresponding 
	to the total population, a subset $C\subseteq P$ of people who believe that Clinton is honest, a subset
	$G\subseteq P$ who believe that Gore is honest, and then the subset who believe that both are honest
	is given by the intersection $C\cap G$. 
	Whether a participant belongs to one of these sets is unaffected by asking if the participant also belongs to another. Asking a question may cause a system to output an answer, but it does not change the system's internal state.
	To account for the changing probabilities observed above, we would need to
	be able to model the question “Is $x$ in $G$?” as partly depending on whether the model has been asked 
	“Is $x$ in $C$?”.  
	
	There are several other known examples that violate the basic classical assumption that
	probability judgments depend on comparing the relative sizes of fixed sets \citep{tversky1983extensional}.
	These include so-called conjunction fallacies and disjunction effects, where the
	probability of a combination of events is considered to be greater or less than what would
	be possible if the probability for each event is measured and the probabilities are then combined \citep{tversky1992disjunction}. 
	
	A well-known example is the Prisoner's Dilemma paradox, 
	in which there are two prisoners with no means of communication, whom the police have reason to 
	believe may be connected to the same crime. Each prisoner is offered two choices: 
	
	\begin{itemize}
		\item `Betray' and attest that the other prisoner is a partner in crime, or, 
		\item `Cooperate' by \emph{not} implicating the other prisoner in the crime.  
	\end{itemize}
	
	If both prisoners cooperate, they both go to jail for 2 years. If one cooperates and the other betrays, then
	the betrayer only goes to jail for 1 year, and the cooperator for 5 years. If both betray one another, both go to jail for 3 years. There are many variants of this setup, studied in psychology and game theory since 
	at least the 1950s \citep{orrell2020quantum} (\S{8.2}).

	From the way the game is setup, even though the smallest jail time overall 
	is when both partners cooperate,
	each participant gets a shorter jail sentence if they betray their partner, 
	whichever choice the partner makes.
	This cold logic is confirmed in many experiments: when told that their partner
	has betrayed them, 82\% of participants betray, and when told that their partner has cooperated,
	72\% of participants betray.
	The paradoxical finding, however, is that the probability of defecting is reduced to 64\% if the
	partner's choice is unknown. (\hl{These} %MDPI: Footnote is not permitted in this journal, so we have moved it into the text, please confirm the whole text. %Confirmed
	are averages across many experiments, summarized by 
	\citet[Table 9.4]{busemeyer2012quantum}.) 
	This is paradoxical because it violates the “Sure Thing Principle”
	of \citet{savage1954foundations}, or in mathematical terms, the classical law of total probability, 
	which in this case would state that:
	\begin{equation}P(B) = P(B|C)P(C) + P(B|C')P(C').
		\label{eq:total_prob}
	\end{equation}

	Substituting in the observed numbers above, if $P(B) = 64\%$ is smaller than both  
	$P(B|C) = 82\%$ and $P(B|C') = 72\%$, and $P(C) + P(C') = 1$, there is 
	no possible value of $P(C)$ and $P(C')$ that could satisfy this equation.
	
	These are two of the many examples from the seminal works of Dan Kahneman and Amos Tversky, amongst others, which have illustrated the systematic bias in judgments, relative to the classical expectation that probabilities should be solely derived from frequencies
	\citep{kahneman1973psychology,tversky1974judgment,tversky1983extensional}. 
	Other well-known examples include conjunction fallacies: for example,
	participants told that a character, Bill, is intelligent  and strong in math, but unimaginative,
	considered it more likely that “Bill is an accountant who plays jazz for a hobby” 
	than just “Bill plays jazz for a hobby” \citep{tversky1983extensional}.

    Sometimes, violations of classical probability laws have common explanations.
    In the Prisoner's Dilemma, participants could be described as giving benefit of doubt to their partners:
    in some situations, not knowing an outcome may encourage someone to view a situation more positively.
    Sometimes the opposite happens, for example when we experience risk aversion.
    Risk aversion is well-known in economics, and has been demonstrated in experiments: for example, 
    in a paper called {\it The uncertainty effect: When a risky prospect is valued less than its worse outcome},
    \citet{gneezy2006uncertainty} reported that:
    
    \begin{quote}
    For example, people are willing to pay an average of \$26 for a \$50 gift certificate, but only \$16 for a lottery that pays either a \$50 or \$100 gift certificate, with equal probability.     
    \end{quote} %Recent addition, for recommended citation.
	
	The analysis of probabilities and the inequalities that must hold between them figures prominently
	in George Boole's {\it {The Laws of Thought}} \citep{boole1854laws} (Ch. 16--21) and is developed
	in Boole's subsequent works, in which inequalities between probabilities are described as 
	`conditions of possible experience'. Boole focused particularly on the issue of compatibility between
	individual (marginal) and joint probabilities: for example, if 6 out of 10 objects are green 
	and 7 out of 10 are square, there must be at least $6 + 7 -10 = 3$ objects that are both green and square.
	With this background, one way to consider the work described here is as an extension of Boole's 
	work on joint and marginal probabilities, to include conditional probabilities of unknown events.
	This topic is already known to be connected with quantum theory, because the famous Bell and CHSH
	inequalities of quantum physics are examples of Boole's conditions, as recognized and explained
	by \citet{pitowsky1994george}. 
	
	%%%%%%%%%
	\section{Quantum Probability Models for Cognitive Behavior}
	\label{sec:qc_intro}
	
	This section describes some examples of the quantum approaches that have been used to model some 
	of the above scenarios more effectively, focusing particularly on order and disjunction effects.
	The field of quantum cognition has developed in cognitive science since the 1990s, and 
	uses mathematical structures from quantum theory as models for human behavior in ways that
	solve precisely the problems posed in the previous section.
	Several authors
	have contributed key insights and results in this field, which is now established enough
	to have comprehensive surveys including those of \citet{busemeyer2012quantum,blutner2016quantum,pothos2022quantum}, and
	in economics, \citet{orrell2020quantum}.
	There are many variants of the approaches introduced, and this section is intended to summarize the main themes
	rather than to provide an exhaustive review.
	Here we explain key aspects of the models for order and 
	disjunction effects that motivate the subsequent quantum circuit implementations.
	
	Common themes used in these approaches include:
	
	\begin{itemize}
		\item The states of systems are represented by vectors.
		\item Outcomes of `measurements' (for example, answers to questions) are modeled 
		by projecting state vectors onto eigenvectors representing different outcomes.
		\item The probability of a particular choice or answer is determined by the
		square of the scalar product
		(hence the angle) between the system's state and the outcome state.
		\item A system can be in a superposition of various different states.
		\item Contributions to these superpositions can interfere (constructively or destructively). 
		This sometimes represents cognitive couplings between events.
		\item Measuring the system forces it to `choose' or collapse into a specific output state. 
	\end{itemize}
	
	A core notion in quantum cognitive models is that the violation of classical probabilities 
	in disjunction effects happens because the states representing the as-yet-unknown outcomes of the events interfere with one another. 
	% for example, in the Prisoner’s Dilemma situation, subjects appear to demonstrate a 'benefit of the doubt' tendency.
	Then, when the uncertainty is resolved (by telling a subject a particular outcome, or sometimes by changing the wording to encourage this kind of analysis), the scenario `collapses', the available options are reduced, and the interference vanishes.
	
	%%%
	\subsection{Example with Order Effects}
	
	Order effects such as the Clinton--Gore example have been described using
	quantum probability as an alternative to classical probability. Quantum probability
	depends on comparing angles rather than volumes, and crucially, measuring a system
	causes it to `collapse' from a superposition of states. The state is
	projected onto whichever pure state is observed, with a probability determined by the 
	magnitude-squared of the projection output.
	As projections do not commute with one another, the order of projections matters:
	so the probability of different outcomes depends on the order of measurement.
	This property has been used successfully to model order effects, including
	work established in the survey {of} 
	\citet{busemeyer2006quantum}. 
	
\begin{figure}[H]
		\includegraphics[width=\linewidth]{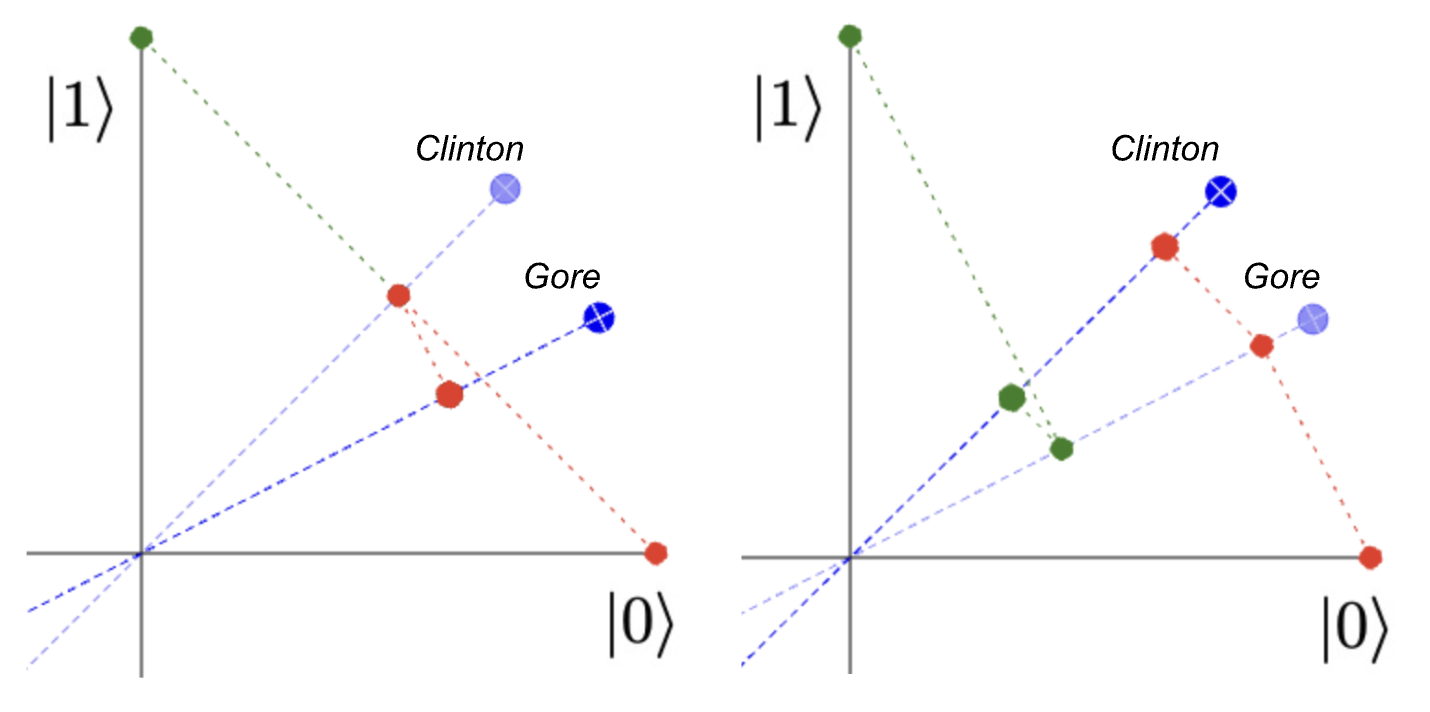}
		\caption{Non-commutative projections model the order effect in the Clinton--Gore scenario. 
			We see that the projection of the $\ket{0}$ vector onto the \textsf{{\it {Clinton}  
			}} axis gives 
			a point further from the origin if we first project onto the \textsf{{\it {Gore}}} axis (\textbf{right}) 
			rather than if we just project the $\ket{0}$ vector onto the \textsf{{\it {Clinton}}} axis (\textbf{left}).
			These diagrams show only the quadrant with positive real coordinates, so if the {\it {Gore}}
			axis is at angle $\theta$ above the horizontal, it appears at the point 
			$\cos(\theta) \ket{0} + \sin(\theta) \ket{1}$. In reality, the
			coordinates can be any complex numbers $\alpha$ and $\beta$ such that $|\alpha^2| + |\beta^2| = 1$.}
		\label{fig:clinton_gore}
	\end{figure}

	These models are often described using images such as those in Figure~\ref{fig:clinton_gore} 
	(for example, see \citep{pothos2022quantum}, p. 757).
	It is 
	clear that the projection of the $\ket{0}$ vector onto the \textsf{\textit{Clinton}} axis is further from the
	origin (hence represents a larger probability) if the projection chain goes via the \textsf{\textit{Gore}}
	(right) axis than when it jumps all the way to \textsf{\textit{Clinton}} at once (left). This follows from
	the double-angle formula $\cos(a + b) = \cos(a)\cos(b) - \sin(a)\sin(b)$ with $0\leq a, b \leq \frac{\pi}{2}$,
	because $0 \leq \sin(a), \sin(b)$, and so $\cos(a + b) \geq \cos(a)\cos(b)$.
	
	Key quantum concepts used here are superposition and projection. The initial $\ket{0}$ state can
	be written as a superposition of states representing the different outcomes: for example, a weighted sum
	of the states representing the beliefs “Clinton is honest” and “Clinton is not honest”.
	In the process of answering a question, the state vector of the
	system is projected or 'collapses' to the state representing the given answer. This is a crucial feature
	of quantum cognitive models: mathematically, the use of chains of projections that change the state of the system
	naturally gives rise to order effects, because projection operators do not commute with one another.
	
	%%%
	\subsection{Example with Disjunction Effects}
	\label{sec:disjunction_example}
	
	Quantum cognitive models use interference to model disjunction effects. 
	The violation of classical probabilities happens because the states representing the as-yet-unknown outcomes of the events interfere with one another,
	including interference between incompatible outcomes. 
	This can raise or lower the probability of an event, sometimes beyond the bounds that would be possible under the
	classical law of total probability (Equation~(\ref{eq:total_prob})).
	For example, in the Prisoner’s Dilemma situation, participants appear to demonstrate a `benefit of doubt' tendency:
	when the first participant's decision is unknown, the second participant's probability of defecting is lower.
	Then, when the uncertainty is resolved (by telling a participant a particular outcome, or sometimes by changing the wording to encourage this kind of analysis), the scenario `collapses', the available options are reduced, and the interference vanishes.
	\citet{busemeyer2012quantum} comment on this as follows:
	\begin{quote}If choice is based on reasons, then the unknown condition has two good reasons.
		Somehow these two good reasons cancel out to produce no reason at \mbox{all!~\citep{busemeyer2006quantum} ({p. 267).}
		}
	\end{quote}
	Similar famous problems introduced by \citet{tversky1992disjunction} 
	include the Vacation Problem (participants are more likely to book a vacation
	if the result of an important exam is known rather than unknown, whether the known result is pass or fail)
	and the Two-Stage Gambling problem (participants are more likely to gamble a second time if they know
	the outcome of a first gamble, whether or not they won or lost).
	
	The mathematics of such models is described in detail by \citet{blutner2016quantum} (\S{4}) and \citet{orrell2020quantum} ({Ch 4}),
	and for a summary see \citet{pothos2022quantum} ({p. 759}). A key part is that the unknown situation ``$A$ may or may not cooperate''
	is written as a disjunction $A\vee A'$, but this is a quantum rather than a classical disjunction. This means 
	that states of the form $\psi = \lambda A + \mu A'$ are contained in $A\vee A'$ even if $\psi$ is contained in neither $A$ nor $A'$
	directly. This is called the {\it {non-distributive}  
	} property of quantum logic and it explains much of the difference
	between classical (Boolean) and quantum (vector) logic \citep{widdows2021quantum}.
	This leads to a more general quantum version of the law of total probability which takes the form:
	\begin{equation*}P(B) = P(B|A)P(A) + P(B|A')P(A') + \partial(A, B)
	\end{equation*}
	where $\partial(A, B) = P(ABA' + A'BA)$. This term is related to a phase factor $\varphi$ introduced in the derived equation 
	$P_v(ABA' + A'BA) = 2\sqrt{P_v(B|A)P_v(A)}\sqrt{P_v(B|A')P_v(A')} \cdot \cos(\varphi)$, 
	where $\varphi$ relates to the impact of knowing $A$ or $A'$ for assessing the likelihood of $B$.
	
	An appropriate phase angle $\varphi$ can be deduced by knowing $P(B|A)$, $P(B|A')$ and $P(B)$ when the outcome of $A$ is unknown.
	As the number of terms increases, there are many more potential correlations, and the number of possible interference terms
	grows exponentially. This is an opportunity and a challenge: we can represent rich and varied correlations, but the number of parameters
	becomes classically unscalable and there is lots of potential redundancy. Heuristic methods for selecting these parameters 
	have been proposed in the context of quantum Bayesian networks \citep{moreira2016quantum}.

	\subsection{Bayesian Theory and the Cognitive Relevance of Classical Logic}
	
	The departure from classical logic and probability in this paper is not novel. 
	Classical logic as such is no longer considered a viable approach to human behavior, and this was demonstrated convincingly,
	before quantum cognition started to take shape in the 1990s.
	It has been generally accepted since the 1960s that human judgments violate classical logical rules, notably since 
	the work of \citet{wason1968reasoning}. Wason's experiments on reasoning demonstrated systematic discrepancies, between conclusions drawn by participants, and the “correct” inference as predicted by the classical “if P then Q” material implication. Psychologists have explored a wide variety of frameworks for alternative theory and models, including heuristics and biases  \citep{kahneman2011thinking}, and neural networks (as an example application, see \citet{kurtz2007divergent}). 
	
	However, for probabilistic reasoning specifically, an influential approach has been Bayesian probability theory. While it is easily recognized that baseline Bayesian theory cannot accommodate the range of relevant behavioral findings, psychology researchers have sought bounded-rational versions of Bayesian theory \citep{costello2014surprisingly,zhu2020bayesian}. The critical point is that the algebraic structure of Bayesian theory is exactly that of classical logic. Indeed, many of the apparent fallacies in probabilistic reasoning are so surprising, because they break seemingly obvious logical constraints (e.g., the conjunction fallacy, as explained in \citep{pothos2022quantum}). 
	
	While classical logic is no longer employed directly in cognitive modeling, its relevance is still current, because of the enduring interest in Bayesian cognitive models. An analogous point applies to classical circuits: classical circuits acquire ‘relevance’ in current theoretical discussions, exactly because they are the most direct way to implement Bayesian cognitive models. The work in this paper suggests that quantum circuits can play a similarly useful role in the implementation of quantum cognitive models.

	\subsection{Quantum Cognition and Physics}
	
	Quantum cognition, so far, has not depended on any physical apparatus such as a human brain 
	being explicitly `quantum mechanical' in its implementation. This point is sometimes confusing,
	especially as such claims {\it {have}} sometimes been made in what might be called 'quantum consciousness' 
	(see \citet{penrose1989emperor}; a survey of these approaches and their 
	differences is given by \citet{atmanspacher2020consciousness}). 
	To the extent that we continue to assume that the human brain is classical (consistently, with much existing evidence, e.g., \citep{litt2006brain}), 
	then such models are more oriented towards artificial intelligence rather than human cognition, though
	some recent findings also offer some indications  that quantum entanglement is a factor in human consciousness
	\citep{kerskens2022experimental}.
	Whether this distinction is meaningful or not will depend on how much the assumed calculations require quantum physical
	information. 
	As argued by \citet{orrell2020quantum}, this concern is habitually avoided 
	in classical mathematics and mechanics---it is normal to use calculus in economics without 
	worrying about whether the use of mathematics invented for ballistics or planetary dynamics
	is valid in economics, or whether the force of gravity really applies---and quantum cognition
	has taken the same pragmatic liberty with quantum mechanics.
	This approach has made sense partly because so much of the mathematics used in quantum theory has
	applications elsewhere, including throughout contemporary artificial intelligence \citep{widdows2021quantum}.
	
	However, another reason for keeping quantum cognition separate from quantum physics, for the first decades of 
	the relevant research,
	has been that using quantum mechanics itself
	to implement operations was not a practical option. This is now changing rapidly:
	the availability of quantum computers has made it possible to implement at least some models
	of the assumed cognitive operations on quantum hardware. We can now explore this opportunity.
	
	% Mathematically, as derived in \citet{} the final, generalized probabilistic quantum interference formula 
	% \begin{equation}
		% P_{B} = \sigma 
		% \end{equation}
	% requires $2^N$ $\theta$ parameters to tune for $N$ binary random variables. 
	
	% %%%
	% \subsection{Brief Survey of Models?}
	% \citep{busemeyer2012quantum}
	% Quantum Dynamical Model
	
	% \citep{busemeyer2006quantum,pothos2009quantum}
	% The Quantum Dynamical Model is the quantum equivalent of the classical Markov model (3) incorporating time evolution through unitary matrix representation (4). 
	
	% \begin{equation}
		% P_{F}(t) = e^{K \cdot t} \cdot P_{I}(0)
		% \end{equation}
	
	% \begin{equation}
		% Q_{F} = U \cdot Q_{i} = e^{-i \cdot H \cdot t} \cdot Q_{I}(0)
		% \end{equation}
	%  In this case, Q\_{F} is the final belief state of \emph{probability amplitudes}

	%%%%%%%%%
	\section{Quantum Computing and Quantum Circuits Introduction}
	\label{sec:qcomp_intro}
	
	Quantum computers have become a practical reality, and while still small and noisy, they have been used to perform
	basic tasks in parts of AI including natural language processing \citep{meichanetzidis2020qnlp,widdows2022near} and
	image classification \citep{fischbacher2020single}. These works have not demonstrated `quantum advantage' in the sense
	of performing useful tasks that would be computationally intractable on classical hardware. However, they have demonstrated several
	proofs-of-concept, with good mathematical reasons for pursuing these further (these include the exponential 
	dimension growth of tensor products, and correspondences between quantum circuits and other structures
	based on category theory). 

	\subsection{Why Bring Quantum Cognition and Quantum Computing Together?}
	
	As the opportunities for using quantum computers grow, quantum cognition is 
	a natural place to look for applications, because it already uses common mathematical structures, and addresses areas
	known to be difficult for classical logic and probability. 
	
	Both quantum cognition and quantum computing are constrained by the computational/probability rules in quantum theory. Existing quantum cognitive models are intended to be at the ‘computational level’ of explanation, using the framework of~\mbox{\citet{marr1982vision}.}
	The computational level concerns the what and the why questions for the system that is studied, that is “what is the goal of the computation, why is it appropriate, and what is the logic of the strategy by which it can be carried out?” \citep{marr1982vision} ({p. 25}).
	Notably, quantum cognitive models address the question of the computational principles which appear to guide behavior. 
	
	Quantum computing and circuits offer insight into how quantum cognitive models could be implemented. In Marr’s terms, they concern the algorithmic level of explanation, which concerns process explanations of the studied system, specifically the representations that are employed by the system and the algorithms that operate on the representations to produce an output from an input. 
	
	With these thoughts in mind, a quantum circuits implementation of a quantum cognitive model has two main possible aims. First, it could serve as an algorithmic/ process proposal for a corresponding quantum cognitive model. Assuming that there are no real quantum mechanical processes in the brain, a putative brain quantum circuit would be epiphenomenal. Second, a quantum vs. classical circuit implementation (of a quantum cognitive model) bears on the question of human vs. artificial intelligence capacity and limits. 
	
	Both these aims would require extensive further work to substantiate---the present work provides the foundation for such work, by offering the first principled proposal for quantum circuits 
	corresponding to a quantum cognitive model; and proof of concept, in the form of the simulations conducted directly on a quantum computer. 
	
	A related reason for applying quantum computers to quantum cognition is the obvious motivation-by-opportunity---“because we can”. 
	A great deal of investment has gone into developing quantum computers, with anticipated application areas including
	materials science (e.g., molecular simulation) and logistics (because of combinatoric complexity). It is to be expected
	that researchers will try to apply these machines to other areas, just as GPUs (Graphics Processing Units) have
	found extensive applications in machine learning, as well as computer graphics. We expect that, during the 2020s,
	quantum computers will be tried in many more application areas. Domains with established techniques 
	that already use quantum mathematical models are naturally a promising place to start.

	\subsection{A Brief Primer on Qubits, Quantum Gates and Quantum Circuits}
	
	This section explains some of the key aspects of quantum computing needed to understand basic quantum circuits.
	The proposal of using a quantum mechanical computer to simulate natural processes was made by Richard Feynman in 1982,
	and by the end of the 1990s, a clear idea of a quantum computer based on quantum bits (qubits) and quantum logic gates had emerged
	\citep{nielsen2002quantum} ({Ch 1}).
	In the early 2020s, most quantum programming still involves assembling qubits and gates explicitly,
	so this introductory section on quantum circuits is like explaining to a reader already familiar with Boolean logic
	how basic classical logic gates are represented and connected to program a digital computer.
	
	We begin by describing individual qubits. A (classical) bit is a two-level classical system, meaning it can be in one of
	two states $\{0,  1\}$. Bits can be represented using different voltages, currents, light intensities, or magnetic polarities,
	though this material implementation is usually immaterial to the mathematical description. 
	A qubit, by contrast, is a two-level {\it {quantum}} system, meaning it can be in a linear superposition of the 
	two states: if the two levels of the system are written $\ket{0}$ and $\ket{1}$, a qubit can be in a more general
	state $\alpha\ket{0} + \beta\ket{1}$. This gives rise to the popular description that a qubit can be in a state 
	`between 0 and 1', but there is more to a qubit than that. A crucial feature of quantum mechanics is that
	complex numbers are used throughout, for reasons including the fact that the function $f(t) = e^{it}$ conserves
	magnitude for real inputs $t$, whereas either the real part $\cos(t)$ or the imaginary part $i\sin(t)$ would lead to fluctuating overall probabilities. Thus, $\alpha$ and $\beta$ are complex numbers, so the space of possible
	values $\alpha\ket{0} + \beta\ket{1}$ has $2+2=4$ real dimensions.
	
\begin{figure}[H]
        \includegraphics[width=5cm]{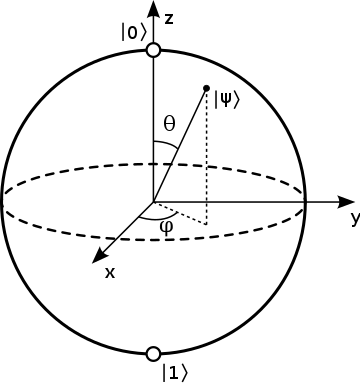}
		\caption{Bloch sphere representation of a qubit. (\hl{From} %MDPI: Please add the access date (format: Date Month Year), e.g., accessed on 1 January 2020.
			\url{https://en.wikipedia.org/wiki/Bloch_sphere}, 
			Creative Commons CC BY-SA 3.0 license. Accessed on 14 March 2023.)} %EditedHere
		\label{fig:bloch_sphere}
	\end{figure}

	The probabilities of observing a qubit in state $\alpha\ket{0} + \beta\ket{1}$ to be in the states $\ket{0}$ or 
    $\ket{1}$ are given by $|\alpha|^2$ and $|\beta|^2$ respectively.
	It follows that $|\alpha|^2 + |\beta|^2 = 1$, so the vector $\alpha\ket{0} + \beta\ket{1}$ has unit length.
    It also follows that the whole state can be multiplied by any unit complex number $e^{i\theta}$ and give the same 
	measurement probabilities (because $|e^{i\theta}z| = |z|$ for any complex number $z$ and real angle $\theta$).
    These two conditions reduce the number of meaningful parameters needed to describe the state of a qubit
	from 4 to 2, and this 2-dimensional state-space is called the {\it Bloch sphere} \citep{nielsen2002quantum} (\S1.2).
	A standard diagram of this structure is shown in Figure \ref{fig:bloch_sphere}.
	The possible transformation operators on a single qubit---called single-qubit gates---are thus isomorphic to the 
	possible rotations of a 2-dimensional sphere, which has 3-dimensions (generated for example by rotations around the $x$, $y$, and $z$-axes). Thus, a quantum bit is not just more expressive than a classical bit: it is more 
	expressive than is suggested just by saying that it can be 'between 0 and 1'.
	
	The potential of quantum computing becomes even more apparent when we consider multiple qubits. A system
	with $n$ qubits might be measured in any of $2^n$ states (all the variants of $\ket{1}\ket{0}\ldots\ket{1}$, etc., which
	would be written $\ket{10\ldots 1})$. Each of these possible states can have its own 'amplitude' or complex coordinate,
	so the number of coordinates needed to describe the system is $2^n$. Thus, every time we add a qubit, we {\it double}
	the number of amplitudes the system can 'hold in memory' (the quotes being a reminder that we cannot observe these 
	coordinates directly, we can only measure different combinations of zeros and ones). Two qubits
	are said to be {\it entangled} when measuring one of them tells us what to expect when measuring the other: for example,
	with a system in the state $\frac{\sqrt{2}}{2}(\ket{00} + \ket{11})$, if its first qubit is measured and found to be in 
	the $\ket{1}$ state, it follows that its second qubit will also be measured in the $\ket{1}$ state.
	Two-qubit gates are vital in quantum computing for bringing about such entanglement. One of the most commonly-used
	two-qubit gates is the Controlled-NOT, or CNOT gate, which performs a complete $X$-rotation (mapping $\ket{0}$ to $\ket{1}$ and vice versa)
	on the target qubit
	if the control qubit is in the $\ket{1}$ state. Other gates can be constructed out of these ingredients: for example, 
	a collection of 3 CNOT gates can be used to make a swap gate that swaps the state of two qubits.

\begin{figure}[H]
		\begin{tabular}{p{3.5cm}p{3cm}p{3cm}p{3cm}}
			\centering
			\includegraphics[width=2cm,trim={1cm -0.6cm 0.8cm 0},clip]{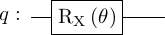} & 
			\includegraphics[width=1.6cm,trim={5.5cm 0 4.2cm 0},clip]{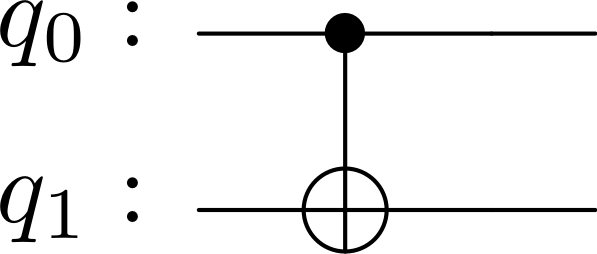} &
			\includegraphics[width=1.3cm,trim={5cm 0 3.7cm 0},clip]{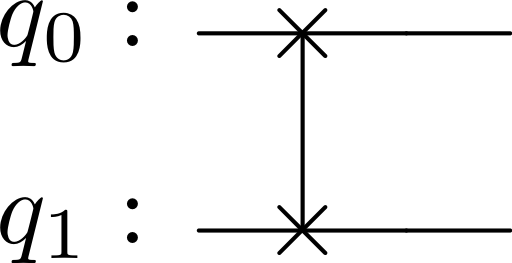} & 
			\includegraphics[width=1.5cm,trim={8cm 1.4cm 4cm 0},clip]{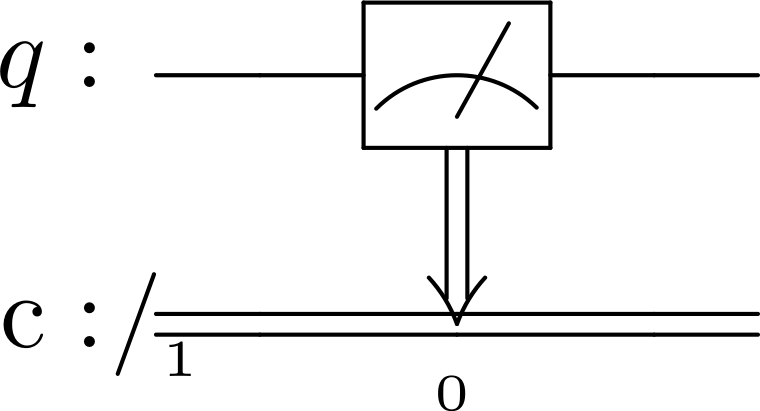} \\
			Single-qubit rotation & CNOT gate & Swap gate & Measurement
		\end{tabular}
		\caption{Basic quantum logic gate diagrams used throughout these examples. A single-qubit rotation gate manipulates the superposition of $\ket{0}$ and $\ket{1}$ states for the qubit. The two-qubit CNOT gate (right) entangles two qubits (the
			top qubit is the control qubit and the bottom is the target qubit).
			The swap gate swaps the states of the two qubits. The measurement operator measures the qubit's value and stores it in the given classical bit.}
		\label{fig:gates}
	\end{figure}

	A {\it {quantum circuit}} is a collection of qubits and gate operations defined in a particular sequence, 
	and typically a {\it {quantum job}} is a computation that proceeds by running a quantum circuit several times
	(each individual run is sometimes called a {\it {shot}}), measuring the results from each run, and combining
	these into a distribution of results that is returned to the user. Figure \ref{fig:gates} shows how 
	the gates and operations used in this paper are depicted in standard quantum circuit diagrams.
	
	The quantum circuits that will be introduced in the next sections, for implementing quantum cognitive models including order and disjunction effects, use standard quantum computing circuit-construction methods: there are no new gates or operations needed that make quantum cognition circuits different from quantum computing circuits in general. However, there are limitations in the current generation of quantum computers (including the lack of mid-circuit measurement), which make some implementation choices more convenient than others today. These decisions are discussed in subsequent sections, as the situations arise.
	
	This is a very brief introduction. Readers who are new to this topic are referred to texts such as those of 
	\citet{busemeyer2012quantum} ({Ch 10}),
	\citet{orrell2020quantum} ({Ch 5}) (specifically focused on quantum cognition), 
	and \citet{bernhardt2019quantum} ({Ch 7}), \citet{nielsen2002quantum} ({Ch 4}) (for quantum
	computing in general). There are also several good online introductions to quantum gates and circuits. (\hl{See, } %MDPI: Footnote is not permitted in this journal, so we have moved it into the text, please confirm the whole text.
	e.g., \url{https://en.wikipedia.org/wiki/Quantum_logic_gate}, \hl{accessed on March 14 2023} 
 %MDPI: Please add the accessed date, e.g., accessed on 1 January 2023. %EditedHere
	and \url{https://qiskit.org/learn/}, \hl{accessed on March 14 2023} c.f. 
	\citep{qiskit2021textbook}.)
	
	%%%
	\section{Quantum Circuits for Order Effects}
	\label{sec:order_circuits}
	
	Having introduced quantum cognitive models and quantum circuits, this section starts 
	to bring these together by introducing a quantum circuit implementation for order effects.
	For this, we return to the example of participants being asked about Clinton or Gore's honesty
	in different orders.
	Each question “Is $X$ honest?” is a `yes/no' question, so can be modeled with a 2-state quantum 
	system, i.e., a single qubit. (\hl{This generalizes the point in Section} \ref{sec:qc_intro} that 
	the $\ket{0}$ state can be written as a superposition of states representing “Clinton is honest” 
	and “Clinton is dishonest”. The $\ket{0}$ state can be written as such a sum for any pair
	of orthogonal vectors $X$ and $X^\prime$, and any such $X$ can be written as a superposition
	of $\ket{0}$ and $\ket{0}^\prime = \ket{1}$.) %MDPI: Footnote is not permitted in this journal, so we have moved it into the text, please confirm the whole text. %Confirmed
	
\begin{figure}[H]
		\includegraphics[width=8cm]{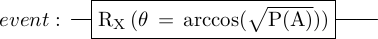}
		\caption{Circuit for setting probability of single event $A$.}
		\label{fig:prob_circuit}
	\end{figure}

	Thus, a quantum circuit for either the Clinton or the Gore question can be created with 
	a single qubit and a single gate: to model one event with two outcomes (say $A$ and not $A = A'$), 
	a qubit is assigned to that 
	event, and a single-qubit rotation is applied to set the appropriate output probability.
	In Figure~\ref{fig:prob_circuit}, an $X$-rotation is used, and the angle $\theta$ is given by the formula 
	$\cos^2(\theta) = P(A) \Longrightarrow \theta = \arccos(\sqrt{P(A)})$.
	
	Since both questions are asked in the same scenario,
	the initial state for both the Gore and Clinton questions can be modeled 
	by a single qubit, while the two questions are represented by measurements along two different axes.
	With such a setup, we can find the appropriate angles $\theta_C$ and $\theta_G$ 
	to model the Clinton and Gore probabilities separately, and if 
	$\theta_G < \theta_C < \pi/2$, we expect that 
	$\cos^2(\theta_G) \cos^2(\theta_C - \theta_G) < \cos^2(\theta_C)$, which means that
	the system is more likely to end up in the $C$ state if it is measured in the $G$ state as a stepping-stone.
	% It is worth noting that this 'short cut' behavior happens mathematically because the scalar
	% product violates the triangle inequality.
	
\begin{figure}[H]
		\includegraphics[width=8cm]{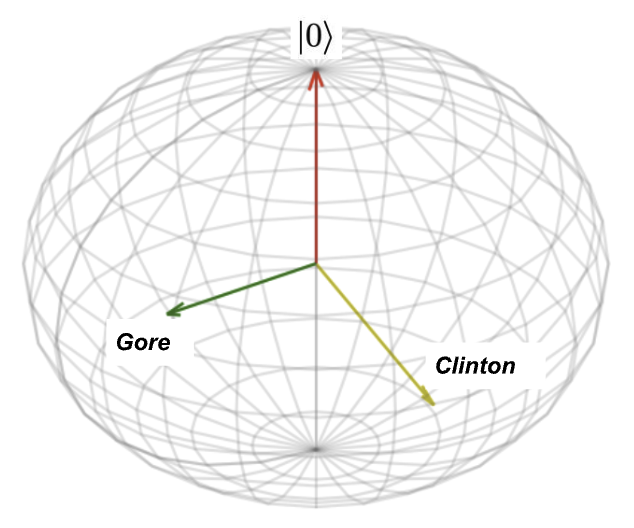}
		\caption{Bloch sphere vectors for Clinton and Gore.}
		\label{fig:cg_bloch}
	\end{figure}

	The 2-dimensional nature of the Bloch sphere arising from the use of complex numbers for coordinates 
	adds a useful degree of freedom here, as seen by comparing Figures \ref{fig:clinton_gore} and \ref{fig:cg_bloch}.
	In Figure \ref{fig:clinton_gore}, each unit vector is described by a single angle, so the whole system comprising of
	the initial vector, the Clinton vector, and the Gore vector, is described by just the two angles $\theta_C$ and $\theta_G$.
	If $P(C) = 0.50$ and $P(G) = 0.68$, this guarantees that $\theta_C\approx 0.785$ and $\theta_G\approx 0.601$, 
	so if the angle between these two is guaranteed to be $\theta_C - \theta_G$, we would infer that 
	$P(G|C) = \cos^2(\theta_G)\cos^2(\theta_C) + \sin^2(\theta_G)\sin^2(\theta_C) \approx 0.668$, which is larger than the
	value of 57\% found in experiments. In this one-dimensional model, Clinton's perceived trustworthiness gets too much of
	a boost by being associated with Gore's.
	
	In the 2-dimensional Bloch sphere model of Figure \ref{fig:cg_bloch}, 
	instead of being predicted by a single angle $\theta$, the Clinton and Gore vectors 
	also have a phase angle $\varphi$, and an appropriate combination of rotations can be used
	to generate any of these states. So as well as fitting the $\theta_C$ and $\theta_G$ parameters
	to give the expected probabilities for $C$ and $G$ on their own, we can also find a parameter $\varphi$
	that fits the expected probability of transitioning from $G$ to $C$, as in Figure \ref{fig:cg_bloch}.
	This enables the model to fit the distance between $C$ and $G$, as well as each of their
	distances from the user's starting point---not just how they are related to the user, but how they are
	related to {\it each other}. If the user is at the pole and the spherical coordinates of the other points are
	$C = (\theta_C, \varphi_C)$, $G = (\theta_G, \varphi_G)$, 
	the symmetries of the sphere (including global phase-invariance)
	ensure that the absolute phase values do not matter---the polar angles $\theta_C$ and $\theta_G$ and the
	difference in phase angle $\varphi_{GC} = \varphi_C - \varphi_G$ are enough to describe the whole scenario.
	This new parameter $\varphi_{GC}$ can be interpreted as the extent to which the user perceives
	the two other items or questions to be related to each other. (The notion that semantic
	vector similarity should take that into account a user's point-of-view was also used
	by \citet{aerts2011similarity}.)
	
	Thus, a single qubit has exactly the right number of coordinates/degrees of freedom to model the order effects
	in two-question scenarios such as the Clinton--Gore example. 
	In general, it is more desirable to have fewer parameters than data degrees of freedom, 
	since exact equivalence undermines the necessity that a model has a particular form. 
	(However, we will see in subsequent sections that for larger circuits, a 1-1 correspondence between
	degrees of freedom in the circuit and parameters of a psychological model is not obvious in higher dimensions.)
	It also preserves an interesting symmetric property whereby
	the gain in trustworthiness for Clinton when compared with Gore is the same as the loss in trustworthiness for Gore 
	when compared with Clinton. This mathematical property is described as `QQ-equality' {by} \citet{wang2014context},
	in whose survey analysis it is found to hold with surprising accuracy.
	
\begin{figure}[H]
		\includegraphics[width=9cm]{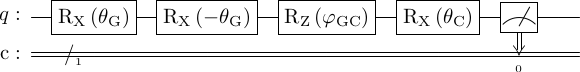}
		
        \small
		\raggedright
		\hl{Gore then} %MDPI: Please check if text in the figure can be moved into figure caption, if yes, please move them into caption %Moving text breaks coupling of text and image. To keep them together, could declare as separate figures, a slightly bigger change but easy to do. Let me know if I should do this. In the meantime, made this text caption-sized.
		Clinton Rotations without Mid-Measurement. (The rotations through $\theta_G$ and $-\theta_G$ cancel each 
		other out in this case: they are there because in principle the state could be measured after the first rotation, as
		simulated with the swap gates in the other two circuits below.)
		
		\vspace{0.2in}
		
		\includegraphics[width=10cm]{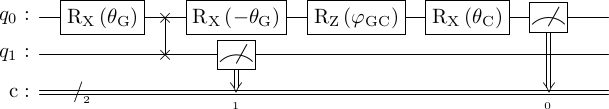}
  
		\hl{Gore} %MDPI: Please check if text in the figure can be moved into figure caption, if yes, please move them into caption %Checked, see above
		then Clinton Rotations with Mid-Measurement of \ket{0}
		
		\vspace{0.2in}
		
		\includegraphics[width=10cm]{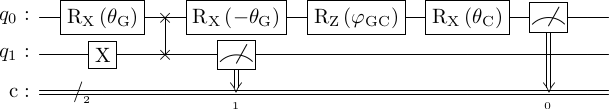}
		
		\hl{Gore} then Clinton Rotations with Mid-Measurement of \ket{1}
		
		\caption{Circuits implementing the Clinton--Gore order effects, 
			with and without mid-measurement.}
		\label{fig:cg_circuits}
	\end{figure}

	The three circuits 
	used to model the scenario and recover the probabilities for the questions “Is Gore honest?” and 
	“Is Clinton honest?”, with and without being asked the Gore question first, are shown in Figure~\ref{fig:cg_circuits}.
	They are slightly more complicated than might be expected because, in practice, there are also constraints 
	that arise from running on contemporary NISQ (Noisy Intermediate Scale Quantum). 
	Some of these choices are described in detail by \citet{nielsen2002quantum} (\S{4.4}), including 
	the {\it Principle of Deferred Measurement}.
	
	Instead of measuring the state $\ket{0}$ in the basis of the {\it {Clinton}} or {\it {Gore}}
	axes, an appropriate rotation is applied and then the state is measured in the usual 
	computational basis. These are mathematically equivalent, and enable 
	measurements along any notional axis to be performed while only physically implementing measurement for one
	axis (typically the $z$-axis). 
	So, to check “Is $A$ honest?” with honest being the $\ket{0}$ state, we set the state to $A$ and check for honesty,
	rather than setting the state to honesty and checking for $A$.
	Thus, instead of measuring along the {\it Gore} axis, we perform an $X$-rotation
	through angle $\theta_G$ and then measure along the standard $z$-axis to see if the qubit is in state $\ket{0}$ or $\ket{1}$. Or, if we choose not to measure this state but only the 
	{\it Clinton} question without the comparative {\it Gore} context, we append the gate operators
	to map from the {\it Gore} state to the {\it Clinton} state and measure that instead. This 
	gives the topmost circuit in Figure \ref{fig:cg_circuits}.
	
	The other two circuits in Figure \ref{fig:cg_circuits} model the situation where the question “Is Gore honest?” is asked first. However,
	measuring the same qubit more than once during a quantum circuit is error-prone. This is sometimes described
	as the {\it {mid-circuit measurement}} problem, and though the recent study of \citet{hua2022exploiting} 
	“discovered non-trivial potential for qubit reuse”,
	this is early-stage and not yet used widely. 
	In our implementation, instead of using one qubit and measuring its state twice, two qubits are used, 
	which can be swapped with one another after the first $\theta_G$ rotation.
	Such a placeholder qubit is often called an {\it {ancilla}} (c.f. `ancillary').
	The final state in the main qubit is the same as the state that would result 
	whenever the state that is measured mid-circuit is the same as the initialized state of the qubit that was swapped in. 
	Hence, the difference between the middle and the bottom circuit in Figure \ref{fig:cg_circuits}:
	the initial full $X$-rotation on the ancilla qubit makes it so that the $\ket{1}$ state is 
	inserted instead of the $\ket{0}$ state, which corresponds to answering the first
	question “Is Gore honest?” with “No”.
	
	There are various ways such a protocol can be implemented as well as swapping qubits: 
	as an alternative solution to the same problem, \citet{orrell2020quantum} (\S{5.5}) proposes the use of two qubits connected by a CNOT gate.
	Such techniques reduce the engineering problem from implementing many reliable gate options, and many
	reliable measurement options, to keeping the same requirements on gate implementation, while reducing the requirements for
	measuring reliably along any axis at any time.
	
	Using these substitutions, the probabilities for what would have happened had the intermediate state been measured can
	be reassembled, but at the cost of using an extra qubit and circuit complexity.  
	Thus, the second and third circuits in Figure \ref{fig:cg_circuits} give the 
	probability of saying that Clinton is honest and Gore is honest/dishonest, depending on whether
	the ancilla qubit $q2$ is $X$-rotated before being swapped for $q1$. 
	The relative proportions
	of the relevant outcomes are then used to reconstruct the conditional probabilities as if the 
	intermediate state was measured. For example, in the bottom circuit of Figure~\ref{fig:cg_circuits},
	the ancilla qubit $q_1$ is swapped with $q_0$ in the state representing the answer “Gore is dishonest”,
	and so the frequency of answering with the combination “Clinton is honest given that Gore is dishonest” is given
	by the conditional probability $P(q_0=0, q_1=1 | q_1=1)$, i.e., the~ratio~$\frac{\#\ket{01}}{\#\ket{01}\ +\ \#\ket{11}}$.
	
	% This is done now. Good to check when possible.
	% \todo{To make sure of:
		% let me say I have done some circuits in the past (i did physics in my first degree) and I have read large parts of N\&C. However, I do not have handy all this knowledge (I do not regularly use it). Given all this, I cannot parse Fig 4, which is crucial for the present proposal. 
		
		% As with other comments, whether you do something about it or not depends on how much you want to engage with beh scientists. 
		
		% I do not think you need to do lots, but explaining at least how the first circuit of Fig 4 works would help (including in terms of what the various operations correspond to). 
		
		% For at least the first circuit, showing the operations algebraically would again help make the text more palatable. 
		% }

	This example is quite simple, but already demonstrates some of the issues and opportunities
	when implementing quantum models from the literature on quantum computers. The use of
	complex numbers and combinations of qubits give rise to opportunities to build models and fit 
	parameters in new ways, and in particular, the extra phase parameter $\varphi_{GC}$ enables us
	to model the relationship between three states (the participant's initial state, and the 
	states representing the answers to two questions) exactly and without redundancy
	for this small model.
	The design choices made when implementing this model using quantum circuits
	also depend on current hardware features, some of which are evolving quite quickly.

	\subsection*{Conditions on Probabilities in Order Effect Circuits}
	
	Quantum probabilities are constrained by an axiomatic framework analogous to that of classical
	probability theory, but with notable differences. Some of these differences depend on 
	choices of implementation for particular operators. For example, the belief “Clinton is honest, and
	Gore is honest” could be represented as an intersection of higher-dimensional subspaces
	representing “Clinton is honest” and “Gore is honest”, respectively. This is like the original quantum
	logic of \citet{birkhoff1936logic}. However, if the beliefs “Clinton is honest” and “Gore is honest” are represented
	just as one-dimensional lines (equivalent to points on the Bloch sphere for an individual qubit), their
	intersection is typically empty or trivial (and even if these concepts are represented as higher-dimensional
	subspaces, the projection onto the intersection of these subspaces is only well-defined when the 
	projections onto each subspace commute with one another \cite{isham1995lectures}, \S{9.2.3}).
	
	Instead, our order effects circuit models use the notion that the probability that a participant 
	says “Clinton is honest” and “Gore is honest” is given by the projection onto the {\it Clinton} state
	followed by projection onto the {\it Gore} state.
	Projecting onto $A$ and then onto $B$ is sometimes called an `$\& \mathrm{then}$' operator, following \citet{busemeyer2006quantum} and \citet{pothos2022quantum}. Conditions associated with this operator include:
	
	\begin{itemize}
		\item In quantum theory, we can have $P(A \andthen B) > P(B)$, because projecting onto
		$A$ and then onto $B$ can give a higher probability than projection directly onto $B$.
		\item However, it has to be the case that 
		$P(A \andthen B) < P(A)$ (so this model does not allow double conjunction fallacies). 
		
		% \item There are constraints to the resolution of the identity analogous to those for Bayesian theory: 
		% \[P(A) + P(A^\prime) = 1\] 
		% and 
		% \[P(A\andthen B) + P(A\andthen B^\prime) 
		% + P(A^\prime\andthen B) + P(A^\prime\andthen B^\prime) = 1.\]
		
		\item Since $P(A \andthen B) \neq P(B \andthen A)$,
		exchanging the orders of $A$ and $B$ terms in these conjunction expressions can lead to different outcomes.
	\end{itemize}
	
	Large differences between $P(A \andthen B)$ and $P(B \andthen A)$ can occur. For example, if the initial
	state is $\ket{0}$ and the $A$ state is $\ket{1}$, then $P(A) = 0$, so also $P(A \andthen B) = 0$. If the 
	$B$ state is $\frac{1}{\sqrt{2}}(\ket{0} + \ket{1})$, half-way inbetween, then the probability of 
	projecting onto $B$ is $\frac{1}{2}$, the probability of projecting from there onto $A$ is also $\frac{1}{2}$, 
	so the combined probability is $\frac{1}{4}$. This property is sometimes referred to as the {\it quarter law} 
	\citep{orrell2020quantum} (\S4.7).
	Other basic axioms and properties of quantum probability are described by \citet{busemeyer2011quantum} and \citet{yukalov2016quantum}.
	
	In classical logic and probability, these differences do not arise: asking “Is the system in state $A$?”
	does not affect the state of the system itself, and modeling a conjunction as an intersection of sets,
	or as two set membership questions asked in succession, amount to the same thing. It follows that 
	any system where $P(A \andthen B) \neq P(B \andthen A)$ is not classical in this sense. This is just a restatement
	of the mathematics behind Heisenberg's Uncertainty Principle: if two operators do not commute with each other,
	then measuring the output of one operator first affects the behavior of the second. 
	(\hl{Since} quantum measurements are inherently probabilistic, 
	establishing empirically that a system is behaving non-classically requires several experiments or shots,
	and statistical analysis to show that differences are significant, which motivates concepts
	such as the {\it {shot noise level}} (SNL) of \citet{ukai2011demonstration}.)  %MDPI: footnote is not allowed in the Journal. we put them in brackets next to a word which to is addressed, please check %Checked

	%%%
	\section{Extending the Order Effects Model with Subjective Bias Activation}
	
	This section describes a useful extension to the order effects circuits
	to include a selective behavior, so that the first question is
	asked only when a given initial condition is met for that participant.
	
\begin{figure}[H]
		\includegraphics[width=10cm]{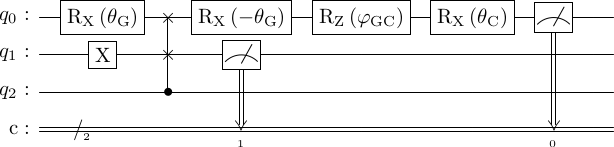}
		\caption{Order effect circuit with an extra qubit $q_2$ that controls
			whether or not the participant is asked the first question.}
		\label{fig:decision_with_activation}
	\end{figure}

	Consider the circuit in Figure \ref{fig:decision_with_activation}. It is just like the bottom circuit
	in Figure \ref{fig:cg_circuits}, except that there is an extra qubit $q_2$ which controls
	the swap gate. In other words, the swap gate (which simulates asking the intermediate question) only operates
	if $q_2$ starts in an excited $\ket{1}$ state, rather than a ground $\ket{0}$ state. This means that the circuit
	will behave differently based on the input state: so if the input state is a model for the participant's initial 
	state or point-of-view, this can be used to model what happens when users start from different states.
	
	Note that as the circuits we consider have more qubits, there are more parameters that could be used but are not.
	For example, in the ancilla qubit $q_1$ and the new priming qubit $q_2$, we have only considered them as being activated or unactivated,
	without considering partial activation and whether the use of complex numbers introduces new opportunities for phase interference effects.
	In theory a state with $n$ qubits can represent $~2^n$ complex parameters (or more strictly, $2^n$ $-$ 2 because normalization and global phase-invariance
	remove 2 degrees of freedom), and as models grow, it is unlikely that we will have an individual role for each parameter. We regard the number of parameters 
	as a computational feature rather than a psychological reality: by analogy, in classical computational modeling, the fact that bits are grouped into
	bytes of 8 bits does not mean that the number of parameters used in classical computational models should always be divisible by 8.
	
	In general, $q_2$ can simulate attentional processes, which determine which questions are considered in a thought process---and so can lead to individual differences.
	A similar design was used by \citet{kvam2017quantum} to model the interaction between user beliefs and user judgments:
	in particular, the extra `belief' qubit is excited when a user recognizes a concept and this cue
	can influence subsequent beliefs or decisions:
	
	\begin{quote}
		If an indeterminate cue
		influences beliefs (via a U-gate operation), evaluation of the cue should affect subsequent evaluation 
		of beliefs and information about the criterion should affect beliefs about the cues \citep{kvam2017quantum}.
	\end{quote}
	
	This can be used to explain why different people respond differently to different cues. The argument can be put differently from the
	perspective of information overload leading to disagreement presented by \citet{pothos2021information}, which 
	considers “well-meaning Alice and Bob debating a complex
	political question” and assumes they “share their questions and outcomes”. This shows that dysfunctional disagreement can 
	still happen if the information is too complicated to consider without simplifying generalizations, and Alice and Bob start with 
	incompatible simplification strategies. Instead, the model in Figure \ref{fig:decision_with_activation} assumes that
	Alice and Bob may have different states for qubit $q_2$, and this guides them concerning which intervening questions to ask:
	even if in theory Alice and Bob `see' the same information, some information will be considered and some will be ignored,
	based upon Alice and Bob's prior state. It is quite possible for this to lead to feedback whereby Alice and Bob continue
	to make different choices about which information they accept and process and which information they ignore: instead
	of information overload, the circuit in Figure \ref{fig:decision_with_activation} models something more like selection bias.
	
	A particular example where this can be seen is in some news headlines: for example, it is more common in the USA for headlines to
	say that a court-case is “a clash between religious freedom and gay rights”, rather than (for example) “a clash between employment law
	and customer service guarantees” because the topics “religious freedom” and “gay rights” are more likely to engage readers than
	`employment law' or `customer service guarantees'. Headlines are deliberately written to connect a story with something the user already cares 
	about, which makes it more likely that the user will read the story. 
	
	The notion of conditional activation can be implemented classically, of course: for example, the representation for each user in the model
	could include a dictionary of configuration parameters, and the implementation could check these parameters and execute an explicit
	\texttt{'if \ldots \ then \ldots'} clause that changes behavior based on these parameters. This approach is more like a user asking
	explicitly “Based upon my prior experiences, am I likely to be engaged by this content?” (which in a quantum circuit would correspond to
	an explicit measurement step). Part of the attraction of the alternative quantum model 
	in Figure \ref{fig:decision_with_activation}
	is that these correlations and choices can be modeled 
	implicitly, which allows for unexpected correlations and behaviors 
	which could be characterized as more instinctive. 
	(In artificial intelligence terms, the classical
	conditional implements a tiny rule-based expert-system, and the quantum circuit implements a tiny quantum neural network.)
	Additionally, in a quantum framework, asking intermediate questions has the potential to alter the underlying mental states, in a specific way, something which does not occur naturally in classical approaches \citep{white2020cost}.
	
	Instead of humans being modeled as rational actors experiencing the same information events and trying to converge to similar beliefs, 
	in this approach, a human's state-of-mind may include a web of activated entangled concepts, which can even be manipulated 
	to make us engage with new information in different ways. This description of what humans are like is very different from the `rational
	actors' often preferred in scientific models since the enlightenment (mid-1600s);
	however, in the 2020s, there is widespread acknowledgment that social and personalized media
	have led to widely divergent self-reinforcing information and belief systems. 
	Despite the discomfort, this day-to-day evidence has made it much easier for scientists to accept that, as humans, we are all biased
	and selective in our approach to information, and that this is something we must take into account rather than try to ignore.
	
	%%%
	\section{Quantum Circuits for Disjunction Effects}
	\label{sec:disjunction_circuits}
	
	Turning to our other main class of cognitive decision models, this section shows 
	how a quantum cognitive model for disjunction effects (as described in Section \ref{sec:disjunction_example} and \citep{busemeyer2012quantum} ({Ch 4}), \citep{orrell2020quantum} ({Ch 4}))
	can be implemented using quantum circuits.
	Common themes of these models include:
	\begin{enumerate}
		\item A method to set the probability of a single event.
		\item A method to connect events saying that the outcome of a particular event may make an output of a subsequent event more or less likely. 
		\item A method to `entangle' events so that states representing different potential events can interfere with one another, including interference between incompatible outcomes.
		\item A method to `measure' events, to model what happens when we learn the outcome of one of the hitherto unknown events and remove the possibility of other outcomes.
	\end{enumerate}
	
	This section demonstrates a combination of basic circuit elements that implement these four key processes.
	
	The process for setting the probability of a single event is just as described above in Figure \ref{fig:prob_circuit}, so we proceed to explain the other three components.

	\subsection{Connecting Dependent Events}
	
\begin{figure}[H]
		\includegraphics[width=12cm]{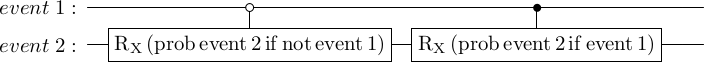}
		\caption{Circuit for setting conditional probability. Note that the white circle means `if this qubit is in state $\ket{0}$'
			and the black circle means 'if this qubit is in state $\ket{1}$'.}
		\label{fig:cond_circuit}
	\end{figure}

	To implement a connection between events that represents the conditional probability $P(B|A)$, 
	we declare qubits $q_A$ and $q_B$ to represent each event.
	We then add a two-qubit gate that implements a partial rotation on the target qubit $q_B$,
	conditioned on the control qubit $q_A$. The angle of this partial rotation is given by 
	the standard transformation from probabilities to angles, $\theta = \arccos(\sqrt{P(B|A}))$.
	The circuit in Figure \ref{fig:cond_circuit} is an example that implements a basic combination 
	of changing the probability of {\it {event 2}}, based on whether {\it {event 1}} happens or not. 
	Similar conditional probability circuits are described by
	\citet{borujeni2021quantum} in the development of quantum Bayesian networks.
	
\begin{figure}[H]
		\includegraphics[width=12cm]{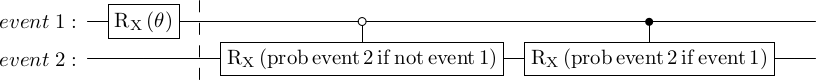}
		\caption{Circuit implementing a simple classical Bayesian network.}
		\label{fig:cond_prob_circuit}
	\end{figure}

	These two components are enough to implement circuits that are equivalent to a classical Bayesian network connecting events 1 and 2, which looks like the circuit in Figure~\ref{fig:cond_prob_circuit}.
	
	This circuit obeys the classical law of total probability, in this case, the rule that
	$P(B)=P(B|A)P(A) + P(B|A')P(A')$.
	The quantum circuit components introduced so far have been demonstrated to give the same outcomes as their classical counterparts. 
	That this is possible with just 2 qubits is striking, but the circuits do not yet model violations of classical rules, 
	such as those seen in disjunction effects.
	
	%%%
	\subsection{Interference between Unknown Outcomes}
	
\begin{figure}[H]
		\includegraphics[width=6.5cm]{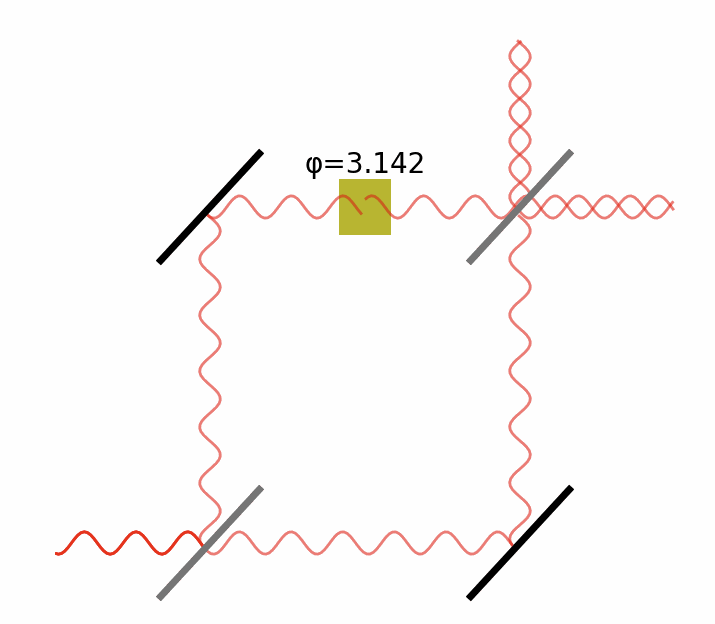}
		\caption{Schematic diagram of a Mach--Zehnder Interferometer.}
		\label{fig:mach_zenhder}
	\end{figure}

	The quantum circuit used to implement interference between unknown outcomes is based on the intuition behind a Mach--Zehnder Interferometer (Figure~\ref{fig:mach_zenhder}). 
	In such a system, a half-mirror splits a beam into two parts, one of those parts may undergo a phase shift, and when the beams are brought back together, they interfere constructively or destructively, based on the phase angle.
	
	\begin{figure}[H]
		\includegraphics[width=5cm]{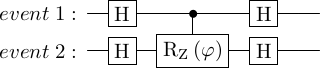}
		\caption{Circuit for simulating interference between unknown outcomes. The 'H' gate is a Hadamard gate which maps the state $\ket{0}$ to a 
			superposition state $\frac{1}{\sqrt{2}}(\ket{0} + \ket{1})$, and $\ket{1}$ to the state $\frac{1}{\sqrt{2}}(\ket{0} - \ket{1}).$}
		\label{fig:interference_circuit}
	\end{figure}
	
\begin{figure}[H]
		\includegraphics[width=\linewidth]{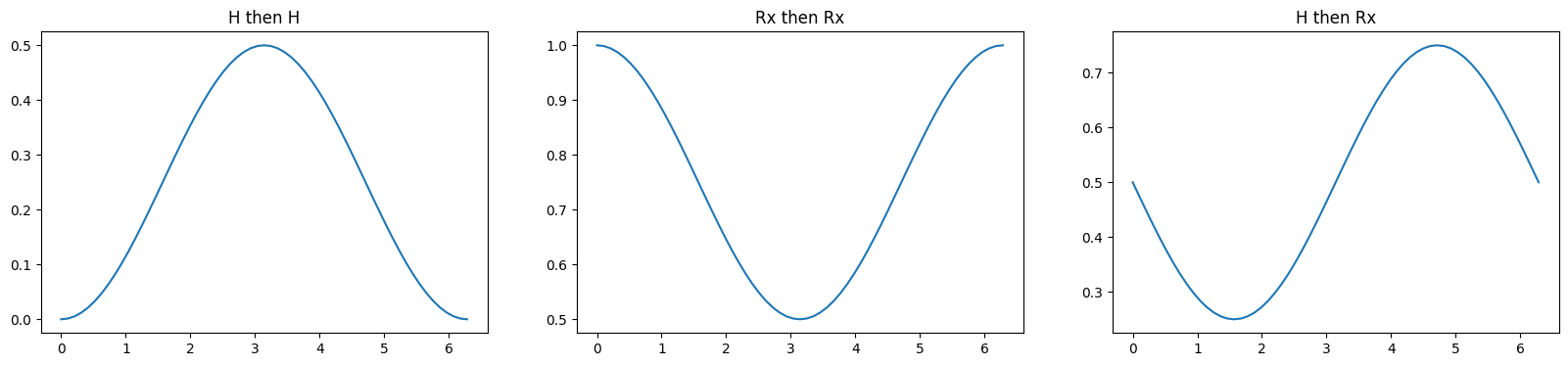}
		\caption{Different output probabilities for the target qubit as a function of the phase angle $\varphi$, when 
			the gates before and after the $R_z(\varphi)$ operation of Figure \ref{fig:interference_circuit} are changed.}
		\label{fig:kernel_varieties}
	\end{figure}

	A quantum circuit that behaves the same way is described by authors including \citet{lee2002quantum}
	and shown in Figure \ref{fig:interference_circuit}. It uses Hadamard gates as the 'beam splitter' that maps a qubit in the state 
	$\ket{0}$ to the state $\frac{1}{\sqrt{2}}(\ket{0} + \ket{1})$, though this is not the only option. 
	For example, changing the single-qubit gates on the target (lower) qubit in Figure \ref{fig:kernel_varieties}
	changes the output probabilities to different ranges. Other operations can be used to shrink the
	window of variation from one-half to smaller ranges. In general, the group of possible two-qubit operations is 
	isomorphic to $U(4)$, the 16-dimensional group of $4\times 4$ complex-valued matrices which conserve probabilities,
	so there are many more degrees of freedom available than just the $\varphi$ parameter in Figure \ref{fig:interference_circuit}.
	On the one hand, this proliferation makes it harder to make direct correspondences between model parameters and 
	psychological factors. On the other, it supports a rich variety of representations. More work is needed to understand
	which families of representations are appropriate for modeling behavioral findings.
	
	% The transformation that occurs by applying the kernel from Figure \ref{fig:interference_circuit} is the following:
	
	% \begin{equation}
		%     \vert 0 \rangle \vert \psi \rangle  \xrightarrow{H} \frac{1}{\sqrt{2}} (\vert 0 \rangle + \vert 1 \rangle)\vert\psi\rangle \xrightarrow{c-U}
		%     \frac{1}{\sqrt{2}} (\vert 0 \rangle + e^{i\varphi}\vert 1 \rangle)\vert\psi\rangle \xrightarrow{H} e^{(i\varphi/2)}(cos \varphi/2 \vert 0 \rangle + i sin \varphi/2 \vert 1 \rangle)|\psi>.
		% \end{equation}
	
	This is not the only option. For example, just applying the gates $H \rightarrow R_z(\varphi) \rightarrow H$ 
	on the same qubit will produce interference between the two states in a single qubit. Instead, using two-qubit 
	configurations to generate interference enables the model to implement an intuition such as 
	“the uncertainty in event 1 affects the phase change in event 2”. 
	This leads to the circuit structure of Figure \ref{fig:interference_circuit}. 
	Various gates could be used instead of the Hadamard (H) gates, which lead to different
	relationships between the phase angle $\varphi$ and the probability for event 2.
	Some of these combinations are shown in Figure \ref{fig:kernel_varieties}.
	The decision to use Hadamard gates 
	is a choice that says “if the probability of event 1 is zero, the probability of event 2 will be 
	between zero and one-half”. (This is before any explicit conditional probabilities, based on knowing the outcome
	of event 1, have been included.)

\begin{figure}[H]
		\includegraphics[width=\linewidth]{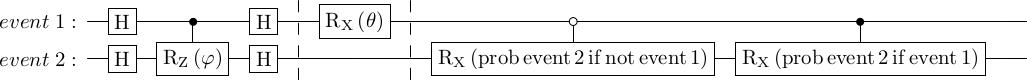}
		\caption{Circuit for conditional probability with interference.}
		\label{fig:kernel_cond_circuit}
	\end{figure}

\begin{figure}[H]
		\includegraphics[width=10cm]{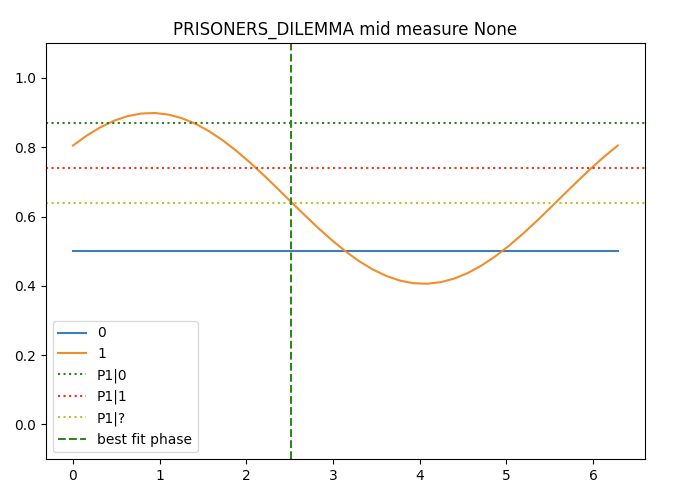}
		\caption{Outcome probabilities for the Prisoner's Dilemma scenario.}
		\label{fig:prisoner_outcomes}
	\end{figure}

	The three components so far are assembled into the following `conditional probability with interference' circuit
	shown in Figure \ref{fig:kernel_cond_circuit}.
	Figure \ref{fig:prisoner_outcomes} shows how this circuit behaves with the numbers filled in from the Prisoner's 
	Dilemma problem
	above. Event 1 is the partner's decision result, with state \ket{0} corresponding to `Betray' and state \ket{1}
	corresponding to `Cooperate'. Event 2 is the participant's decision, with the same correspondences.
	
	The angles represent the average probabilities reported above by \citet{busemeyer2012quantum} (Ch 9). 
	For the probability of event 1 itself (the partner defects), a value of 50\% is commonly 
	used, reflecting the fact that the participants are not given any prior estimate of this event.
	
	The key finding is that the probability of event 2 (the participant defects) occurring varies with the phase angle $\varphi$,
	and it is possible to find a value of $\varphi$ for which the estimated probability is the same as that observed in experiments.
	
	% \todo{note free parameters in the construction 
		% that is how most people would evaluate a model}
	
	%%%
	\subsection{Measuring an Outcome and Removing Interference}
	
	The three components assembled so far can generate the expected probability of event 2 if the outcome of event 1 is unknown. 
	The remaining question is what happens when the outcome of event 1 (such as partner betrays/cooperates)
	becomes known. 
	In quantum theory, this corresponds to measuring the qubit representing event 1, at which point it collapses to 
	the pure \ket{0} or \ket{1} state.
	As in Section \ref{sec:order_circuits}, 
	this can sometimes be modeled without mid-circuit measurement using swap gates and ancilla qubits. 
	
\begin{figure}[H]
		\includegraphics[width=6cm]{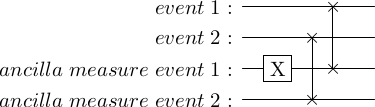}
		\caption{Swapping in ancillas as a proxy for mid-circuit measurement.}
		\label{fig:swaps}
	\end{figure}

	An example is shown in Figure \ref{fig:swaps}.
	In this case, the $X$ rotation on the “ancilla measure event 1” qubit is used to simulate measuring a 
	$\ket{1}$ rather than a $\ket{0}$ for the first event. 
	Swapping in a $\ket{0}$ qubit for event 2 corresponds just to resetting this qubit. 
	
\begin{figure}[H]
		\includegraphics[width=\linewidth]{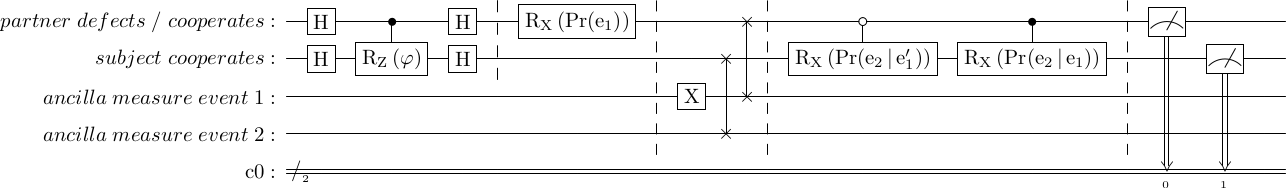}
		\caption{Complete circuit simulating the Prisoner's Dilemma scenario.}
		\label{fig:pd_circuit}
	\end{figure}

	This component is added between the interference component and the conditional probability component discussed above, giving the final Prisoner's Dilemma circuit in Figure~\ref{fig:pd_circuit}.
	This circuit recreates all the desired outcomes for the Prisoner's Dilemma problem. The same circuit structure with different parameter values can be shown to recreate other well-known disjunction problems, 
	including the Two-Stage Gamble and Hawaii Vacation problems studied since their introduction by \citet{tversky1992disjunction}.
	
	While each of the four circuit components used here can be found in different parts of the quantum information 
	literature, we believe that this combination of components is novel, that it accurately models disjunction 
	interference effects from cognitive science, and, to the best of our knowledge, it is the first quantum 
	circuit to achieve this.
	
	% \todo{this is great, and important
		% but you need to give readers some indication of the generative value of this approach 
		% what new, useful thing can we imagine, starting from a circuit-based analysis like the above? }
	
	%%%%
	\section{Developing and Running on Quantum Hardware}
	\label{sec:hardware-sec}
	
	% \todo{you actually ran this on a quantum computer??
		% this would be impressive! 
		
		% however, I think readers will fail to grasp the significance of this 
		% for example, can we say something about the relative time or complexity of resolving the PD problem with a quantum circuit vs. a classical circuit?}
	
	Since all of the circuits in this paper use 4 qubits or fewer, examples 
	of the circuits from Figure~\ref{fig:cg_circuits} for order effects and Figure~\ref{fig:pd_circuit}
	for disjunction effects could be run easily in the regular duty cycle of an
	11 qubit machine made and maintained by IonQ \citep{wright2019benchmarking}. 
	The qubits in this machine are made of Ytterbium ions ($^{171}\mathrm{Yb}^+$), 
	whose ground and excited energy states are distinguished by the hyperfine structure 
	interactions between the spin of the nucleus and the angular momentum of the 
	atom \citep{herzberg1944atomic} (Ch V). The single-qubit and \hl{two-qubit} gates
	are controlled by Raman laser pulses, and the accuracy of these operations is reported as
	99.5\% and 97.5\%, respectively.
	Each circuit was run with 10,000 shots (repetitions),
	which is a key consideration in quantum computing because the 
	output is classical measurement results, not internal quantum states. These measurements are 
	probabilistic by nature, so normally a non-trivial sample collected over many different experiments or shots is necessary.
	
	The circuits for the Prisoner's Dilemma (three versions: partner betrays, partner cooperates, 
	partner decision unknown) produced the expected probabilities (82\%, 73\%, 64\%) very accurately,
	with the largest discrepancy being just 0.37\%. The circuits for the Clinton--Gore order showed
	more variation: the probability of predicting
	“Clinton is honest” with and without being asked the same question about Gore tended to be 
	within 2--3\% of their expected values. This is close to the reported \hl{two-qubit} %MDPI Please choose “two-qubit” or “two-qubit” and apply consistently %Edited to two-qubit throughout.
	gate fidelity for the circuits with
	swap gates, so circuit-specific sources of error were not investigated separately.
	
	These results do not demonstrate a `quantum advantage' in the sense of running something that would
	be intractable on a classical computer. Instead, they demonstrate that we can start with the intended
	behavior of a small psychological model, and implement small quantum circuits that demonstrate
	key statistical outcomes of these models. The present results have some immediate interest, partly because it 
	is sometimes surprising how much can be represented with just a few qubits and gates. (By contrast,
	a circuit using classical bits would require several times more bits and polynomially more gate operations, but these 
	are so much simpler and easier to build using current technologies that a detailed algorithmic comparison 
	would be actively misleading.)
	
	More exciting opportunities here will arise as these components are connected into larger networks. 
	Rather than making the initial experiments here obsolete, this could heighten the value of 
	small software components used to make complicated quantum circuits, because programming and testing
	can be conducted by arranging these higher-level components, rather than addressing individual qubits and gates.
	
	%%%
	\section{Related and Further Work}
	\label{sec:related}
	
	This paper has combined work in quantum circuits and quantum cognition.
	Several of the quantum circuit ingredients here have been used before: for example, the quantum circuit used for interference
	effects (Figure \ref{fig:interference_circuit}) was introduced as a general `quantum Rosetta Stone' by \citet{lee2002quantum}.
	It makes sense to try and assemble larger quantum circuits out of such components to produce higher-level
	structures such as the belief-activation circuits of \citet{kvam2017quantum} and the quantum Bayesian networks
	of \mbox{\citet{borujeni2021quantum}}. The availability of real quantum hardware has encouraged such models to be run on quantum
	computers in related areas, such as natural language processing \citep{lorenz2021qnlp,widdows2022near}, and it is
	natural to expect similar progress in quantum cognition over the next few years.
	
	Quantum cognition has itself been developing the use of corresponding quantum models for many 
	years \citep{busemeyer2006quantum,pothos2009quantum}. The use of complex numbers and the exponential growth in 
	available parameters give more options for parameter fitting, and more challenges in interpreting and using these parameters,
	a situation previously studied by \citet{moreira2016quantum} with heuristic approaches to parameter selection.
	
	Further work should include adapting these techniques to larger systems and datasets. Even the small 
	components discussed here could be rearranged in several ways: for example, the output of a phase interference kernel
	(\hl{Figure} %MDPI: Newly added  information. Please confirm. %Confirmed
	\ref{fig:interference_circuit}) could be used as the input state that may or may not activate a subjective bias 
	entanglement (Figure \ref{fig:decision_with_activation}).
	
	It is possible that such properties could be employed for cognitive findings which have previously benefited from quantum models using entanglement,
	such as conceptual combination \citep{aerts2005theory,bruza2015probabilistic} or bipartite games, such as Prisoner’s Dilemma, with communication~\citep{waddup2021sensitivity}. 
	A particularly interesting direction for application is whether quantum circuits can offer new perspectives on more efficient memory architectures, 
	taking advantage of superpositions and entanglement to multiply query memory representations. 
	
	%%%%%%%%
	\section{Conclusions}
	
	Quantum cognition is a well-established theoretical field that has demonstrated good overlap with real
	world situations, qualitatively and quantitatively, that violate constraints that
	would apply if the decision options are considered separately and their likelihoods inferred using classical probability laws.
	The mathematics of these models (particularly linear algebra)
	is used independently of quantum mechanics, just as the mathematics developed for
	classical mechanics (particularly calculus) has long been used in social sciences.
	Before 2020, this research program was developed
	largely independently of quantum computing, but we are 
	now able to run some of these models on quantum computers for the first time.
	
	Implementing such models on quantum computers is often not straightforward, partly because the approach to computing
	is unfamiliar to most programmers, partly because sometimes alternatives need to be chosen based on a knowledge
	of current hardware support and limitations. One of the main methodological differences between classical and quantum computing
	is that even a single quantum bit (qubit) provides a much richer structure than a classical bit and, as qubits are added,
	this richness grows exponentially.
	
	Quantum programming provides an opportunity to consider new approaches to programming and modeling. 
	In classical computing, at least in theory, all states are known and all operations are predictable. 
	In quantum computing, a state can be unknown and is sometimes deliberately put into a probabilistic 
	state, and deliberately not measured or scrutinized while it interacts with the rest of the system. 
	This encourages different ways of thinking about state and configuration that are intriguing and unfamiliar:
	instead of listing rules to follow or functions to execute, quantum programming is more like preparing
	a collection of states and correlations, letting them evolve, and then probing some part of the eventual state
	with prepared `yes--no' questions. Writing a quantum program as a design process has a lot more in common with 
	designing a complicated psychological survey than classical programming. This is not in itself evidence that
	human thought is a kind of quantum computation; however, it may encourage quantum programmers
	to think about computation not only as a rigid set of steps to follow, but also as a way of preparing systems
	with a network of complicated correlations, which sometimes combine to give surprising results.
	
	This paper has provided some initial quantum implementations, showing some ways in which quantum
	computers lend themselves naturally to these tasks and revealing some areas where hardware constraints
	still constrain design choices. 
	We hope that more research in this area will lead to the establishment of common design patterns that are 
	known to represent some of these non-classical aspects of human cognition naturally and accurately.

\section*{Funding}

EMP was supported by European Office of Aerospace Research and Development (EOARD) grant FA8655-23-1-7220.
The other authors were supported directly by IonQ, Inc.

\bibliography{ionq}

\ifarxiv
    \bibliographystyle{apalike}
\fi

\end{document}